%% file: main.tex
\documentclass[acmsmall,screen,nonacm]{acmart}
\usepackage[utf8]{inputenc}
\usepackage[english]{babel}

\usepackage{enumerate}
\usepackage{hyperref}
\usepackage{graphicx}
\usepackage{csquotes}
\usepackage{cleveref}
\usepackage{mdframed}
\usepackage{xcolor}
\usepackage{xspace}
\usepackage{longtable}
\usepackage{lscape}
\usepackage{rotating}

\acmJournal{TOSEM}

\newcommand{\eg}{e.g.,\xspace}

\newcommand{\viz}{viz.,\xspace}

\title{Smells and Refactorings for Microservices Security: A Multivocal Literature Review}
\author{Francisco Ponce}
\email{francisco.ponceme@sansano.usm.cl}
\affiliation{
    \institution{Universidad T\'ecnica Federico Santa Mar\'ia}
    \city{Valparaiso}
    \country{Chile}
}

\author{Jacopo Soldani}
\email{jacopo.soldani@unipi.it}
\affiliation{
    \institution{University of Pisa}
    \city{Pisa}
    \country{Italy}
}

\author{Hern\'an Astudillo}
\email{hernan@inf.utfsm.cl}
\affiliation{
    \institution{Universidad T\'ecnica Federico Santa Mar\'ia}
    \city{Valparaiso}
    \country{Chile}
}

\author{Antonio Brogi}
\email{antonio.brogi@unipi.it}
\affiliation{
    \institution{University of Pisa}
    \city{Pisa}
    \country{Italy}
}

\date{April 2021}

\begin{abstract}

\textit{Context:} Securing microservice-based applications is crucial, as many IT companies are delivering their businesses through microservices. 
If security \enquote{smells} affect microservice-based applications, they can possibly suffer from security leaks and need to be refactored to mitigate the effects of security smells therein.

\smallskip \noindent
\textit{Objective:} As the currently available knowledge on securing microservices is scattered across different pieces of white and grey literature, our objective here is to distill well-known smells for securing microservices, together with the refactorings enabling to mitigate the effects of such smells.

\smallskip \noindent
\textit{Method:} To capture the state of the art and practice in securing microservices, we conducted a multivocal review of the existing white and grey literature on the topic. 
We systematically analyzed 58 studies published from 2014 until the end of 2020.

\smallskip \noindent
\textit{Results:} Ten bad smells for securing microservices are identified, which we organized in a taxonomy, associating each smell with the security properties it may violate and the refactorings enabling to mitigate its effects. 

\smallskip \noindent
\textit{Conclusions:} The security smells and the corresponding refactorings have pragmatic value for practitioners, who can exploit them in their daily work on securing microservices. 
They also serve as a starting point for researchers wishing to establish new research directions on securing microservices.

\end{abstract}

\begin{document}

\maketitle

\section{Introduction}
\label{sec:intro}
\input{src/1-intro}

\smallskip \noindent
The rest of this article is organised as follows.
\Cref{sec:research-design} illustrates the process enacted to select the reference studies.
\Cref{sec:smells-refactorings} presents the taxonomy of security smells and describes them in detail, together with the refactorings enabling to mitigate its effects.
\Cref{sec:ttv} discuss the potential threats to the validity of our study and how we dealt with them.
 
\section{Search for Studies}
\label{sec:research-design}
\input{src/2-research-design}


\section{Microservices Security Smells and Refactorings}
\label{sec:smells-refactorings}
\input{src/3-smells-refactorings}


\section{Threats to Validity}
\label{sec:ttv}
\input{src/4-ttv}

\section{Related Work}
\label{sec:related}
\input{src/5-related}

\section{Conclusions}
\label{sec:conclusions}
\input{src/6-conclusions}

\bibliographystyle{ACM-Reference-Format}
\bibliography{mybibliography}

\noindent 
{\footnotesize
All links were last followed on \today.
}

\end{document}

%% file: src/1-intro.tex
Microservices are on the rise for architecting enterprise applications nowadays, with big players in IT (\eg Amazon, Netflix, Spotify, and Twitter) already delivering their core businesses through microservices \cite{Thones2015_Microservices}.
This is mainly because microservice-based applications are cloud-native, thus better exploiting the potentials of cloud hosting, and since they fully twin with DevOps and continuous delivery practices \cite{Balalaie2016_MicroservicesDevOps}.
Microservices also bring various other advantages, such as ease of deployment, resilience, and scalability \cite{50-newman-building-microservices}.
Together with their gains, however, microservices bring also some pains, and securing microservice-based applications is certainly one of those \cite{Soldani2018_MicroservicesPainsGains}.

Microservice-based applications are essentially service-oriented applications adhering to an extended set of design principles \cite{Zimmermann2017}, \eg shaping services around business concepts, decentralisation, and ensuring the independent deployability and horizontal scalability of microservices, among others. 
Such additional principles make microservice-based applications not only service-oriented, but also highly distributed and dynamic.
As a result, other than the classical security issues and best-practices for service-oriented applications, microservices bring new security challenges \cite{Soldani2018_MicroservicesPainsGains}.
For instance, being much more distributed than traditional service-oriented applications, microservice-based application expose more endpoints, thus enlarging the surface prone to security attacks \cite{53-lea-microservice-security}.
It is also crucial to establish trust among the microservices forming an application and to manage distributed secrets, whereas these concerns are of much less interest in traditional web services or monolithic applications \cite{44-yarigina-security-challenges}.
Another example follows from the many communications occurring among the microservices forming an application, which ---if not properly handled--- can result in message data being intercepted and in malicious users inferring business operations from such data \cite{Yu2019_MicroservicesFogSurvey}.

Whereas there exist quite much literature on securing microservices, different pieces of literature deal with different aspects of security.
As a result, the currently available information on securing microservices is scattered among a considerable amount of books, blog posts, research papers, videos, and whitepapers.
This hampers consulting the body of knowledge on the topic, both for academic researchers wishing to delineate novel research directions and solutions for securing microservices, and for practitioners daily needing to secure microservice-based applications.
To help both researchers and practitioners, this paper tries to organise the scattered knowledge on securing microservices, by answering to the following two research questions:
\begin{quote}
    (RQ1) \textit{Are there well-recognised smells indicating possible security violations in microservice-based applications?}
    \\
    (RQ2) \textit{How to refactor microservice-based applications to mitigate the effects of security smells therein?}
\end{quote}

An architectural smell can be observed in the architecture of an application, and it is the effect of a bad decision (though often unintentional) while architecting the application, which may negatively impact on the overall system quality \cite{Garcia2009}. 
The effects of architectural smells can be mitigated by refactoring the internal structure of an application or the functionalities of its internal components, while at the same time not changing the functionalities it offers to its external clients.
Even if applying a refactoring requires some development efforts, it is known that refactorings mitigating the effects of architectural smells will most probably improve the overall system quality \cite{Garcia2009}.

To answer to our research questions, we systematically analyzed the available literature on securing microservices to elicit well-known smells, which may possibly result in security violations in microservice-based applications.
We also elicited the refactorings that are known to mitigate the effects of security smells.
In particular, following the recommendations by Garousi et al. \cite{Garousi2016_MultivocalReviews}, we captured both the state of the art and the state of practice in the field by conducting a multivocal review of the existing literature.
We analysed both white literature (\viz peer-reviewed papers) and grey literature (\viz blog posts, industrial whitepapers, books, and videos).
We carefully selected 58 studies published since 2014 (when microservices were first introduced in Lewis and Fowler's blog post \cite{Fowler2014_Microservices}) until the end of 2020, which we then systematically analysed by following the guidelines for conducting systematic reviews \cite{Garousi2019_GuidelinesMultivocalReview,Petersen2008}.
As a result, we obtained a taxonomy that organises the identified security smells together with all refactorings that should be enacted to mitigate their effects.
Finally, we exploited the taxonomy to classify the selected studies, in order to distill the actual recognition of the identified smells and of their corresponding refactorings.

In this paper, we illustrate the results of our multivocal review. 
We first present the taxonomy of security smells, including ten security smells and ten refactorings, organised around three security properties defined in the ISO/IEC 25010 standard for software quality \cite{ISO-standard}, \viz integrity, authenticity, and confidentiality. Then we discuss each security smell by illustrating why it can possibly violate the security property it is associated with.
For each smell, we also discuss the corresponding refactorings recommended by practitioners, by also explaining how such refactoring enables mitigating its effects.

We believe that the results of our study can provide benefits to both researchers and practitioners interested in microservices.
A systematic presentation of the state of the art and practice on well-known security smells for microservices provides a body of knowledge to develop new techniques and solutions, to investigate and experiment research implications, and to set future research directions.
At the same time, it can help practitioners to better understand the currently most recognised security smells for microservices, and to enact the corresponding refactorings to mitigate their effects. 
This is of pragmatic value for practitioners, who can use our study as a starting point for securing their microservice-based applications, as well as a reference in their day-by-day work with microservices.

%% file: src/2-research-design.tex
The objective of this multivocal review is to identify the possible security smells for microservice-based applications, and the refactorings enabling to mitigate their effects.
With the objective of capturing the state of the art and practice in the field, we searched for both white literature (\viz peer-reviewed journal and conference articles) and grey literature (\viz blog posts, industrial whitepapers, books, and videos), in line with what recommended by Garousi et al.~\cite{Garousi2016_MultivocalReviews}.

The structuring of the search string was done by following the guidelines provided by Petersen et al.~\cite{Petersen2008}.
We identified the search string guided by the PICO terms of our research problem, with search keywords taken from each aspect of our research problem.
We anyhow decided ---differently from what indicated by Petersen et al.~\cite{Petersen2008}--- to not restrict our focus to specific research settings.
If restricted to given types of research settings, the results of our analysis could have been biased or incomplete, since some security smells or refactorings might have been over-represented or under-represented in certain types of study.

As a result, our search string was formed by the following terms:
\begin{center}
\ttfamily \footnotesize
    (microservice*)
    $\wedge$
    (security*)
    $\wedge$ \\
    (smell* $\vee$ antipattern* $\vee$ bad practice* $\vee$ pitfall* $\vee$ refactor* $\vee$ reengineer* $\vee$ restructure*)
\end{center}
(where `\texttt{*}' matches lexically related terms).
The search was restricted to studies published since the beginning of 2014 (when microservices were first proposed by Lewis and Fowler~\cite{Fowler2014_Microservices}) until the end of 2020. 

The search of white literature was carried out in the following indexing databases: ACM Digital Library, DBLP, EI Compendex, Google Scholar, IEEE Xplore, INSPEC, ISI Web of Science, Science Direct, and SpringerLink. 
Given the recency of microservice-related studies and the well-known concerns with indexing, Google Scholar played a key role for the initial selection before the inclusion and exclusion stage.
The search for grey literature was instead carried out in the following search engines: Google, Bing, and DuckDuckGo.
Given the amount of results returned by the different combinations of the keywords in the search string (often, hundreds of thousands), and given that each search was repeated on each considered search engine, we adopted the effort bounded stopping criteria suggested by Garousi et al. \cite{Garousi2019_GuidelinesMultivocalReview}.
In particular, for each search on each search engine, we considered the top 250 search hits (\eg the first 25 pages of results returned by the Google search engine).

The above described search criteria were matched by 136 studies, which we carefully screened to keep only those studies that were satisfying at least one of the following inclusion criteria:
\begin{itemize}
    \item A study is selected if it presents {\em at least one security smell} possibly resulting in a violation of a security property defined by the ISO/IEC 25010 software quality standard \cite{ISO-standard}, such as confidentiality, integrity, and authenticity.
    \item A study is selected if it presents {\em at least one refactoring} for mitigating the effects of a security smell, even if the latter is not explicitly mentioned.
\end{itemize}
The inclusion criteria were defined with the ultimate goal of selecting representative studies, discussing the security smells for microservices or their corresponding refactorings.
In addition, we decided to keep only one instance of each representative study, in case they were replicated across multiple pieces of literature, as this often happen in the case of grey literature (\eg we removed \cite{41-gardner-security-microservices} and \cite{11-Raible-Security-Patterns-for-Microservice-Architectures} from consideration, as they were replicating \cite{03-Gardner-Security-in-the-Microservices-Paradigm} and \cite{59-raible-11-patterns-security-microservices}, respectively).

\input{tables/selected-studies}

As a result, 58 studies were selected to be analysed further. 
The selected studies are listed in \Cref{tab:selected-studies} by providing a reference to their full bibliographic information available in the references listed at the end of this article.
The table also classifies each selected study by publication year, colour, and type.

%% file: tables/selected-studies.tex
\begin{table}
\caption{References, publication years, colours (\viz \textit{white} or \textit{grey} literature), and types (\viz \textit{blog post}, \textit{book}, \textit{book chapter}, \textit{conference} paper, \textit{journal} article, or \textit{video}) of the selected studies.}
\label{tab:selected-studies}
\footnotesize
\centering
\begin{minipage}{.34\textwidth}
    \centering
    \begin{tabular}{c@{\hspace{.3cm}}c@{\hspace{.3cm}}c@{\hspace{.3cm}}c}
        \hline
        \textbf{Ref.} & \textbf{Year} & \textbf{Colour} & \textbf{Type} \\
        \hline
        \cite{27-Abasi-Securing-modern-API-microservice}
            & 2019
            & grey
            & blog post     \\
        \cite{34-behrens-starting-avalanche}
            & 2017
            & grey
            & blog post     \\
        \cite{07-Boersma-Top-10-security-traps-to-avoid}
            & 2019
            & grey            
            & blog post     \\
        \cite{51-bogner-microservices-industry}
            & 2019 
            & white
            & conference    \\
        \cite{30-budko-5-things-api-security}
            & 2018
            & grey
            & blog post     \\
        \cite{08-Carnell-Securing-your-microservices}
            & 2017          
            & grey            
            & book          \\
        \cite{10-Chandramouli-Security-Strategies-for-Microservices}            
            & 2019          
            & grey            
            & whitepaper    \\
        \cite{20-da-Silva-Best-Practices-to-Protect-Your-Microservices-Architecture}            
            & 2017
            & grey
            & blog post     \\
        \cite{28-Doerfeld-Control-user-identity-microservices}
            & 2018
            & grey
            & blog post     \\
        \cite{32-douglas-microservice-authentication}
            & 2018
            & grey
            & blog post     \\
        \cite{12-Edureka-Best-Practices-To-Secure-Microservices}            
            & 2019          
            & grey            
            & video         \\
        \cite{45-esposito-challenges-cloud-microservices}
            & 2016
            & white
            & journal       \\
        \cite{18-Feitosa-Microservice-Patterns-and-Best-Practices}
            & 2018          
            & grey            
            & book          \\
        \cite{03-Gardner-Security-in-the-Microservices-Paradigm}            
            & 2017          
            & grey            
            & blog post     \\
        \cite{38-gebel-securing-apis-microservices}
            & 2018
            & grey
            & video         \\
        \cite{29-Gupta-security-strategies-devops}
            & 2018
            & grey
            & blog post     \\
        \cite{17-Hofmann-Microservices-Best-Practices-for-Java}
            & 2016          
            & grey
            & book          \\
        \cite{46-indrasiri-microservices-security-fundamentals}
            & 2018
            & white
            & book chapter  \\
        \cite{48-jackson-go-microservices}
            & 2017
            & grey
            & book          \\
        \cite{15-Jain-Top-10-Security-Best-Practices}            
            & 2018         
            & grey            
            & video         \\
            \hline
    \end{tabular}
 \end{minipage}
 \begin{minipage}{.32\textwidth}
    \centering
    \begin{tabular}{c@{\hspace{.3cm}}c@{\hspace{.3cm}}c@{\hspace{.3cm}}c}
        \hline
        \textbf{Ref.} & \textbf{Year} & \textbf{Colour} & \textbf{Type} \\
            \hline
        \cite{60-kamaruzzaman-microservice-architecture}
            & 2020
            & grey
            & blog post     \\
        \cite{23-Kanjilal-4-fundamental-microservices-security-best-practices}          
            & 2020         
            & grey            
            & blog post     \\
        \cite{19-Khan-How-to-Secure-Your-Microservices}
            & 2018          
            & grey            
            & blog post     \\
        \cite{05-Krishnamurthy-Transition-to-Microservice-Architecture}             
            & 2018      
            & grey        
            & blog post     \\
        \cite{53-lea-microservice-security}
            & 2015
            & grey
            & blog post     \\
        \cite{40-lemos-app-security}
            & 2019
            & grey
            & blog post     \\
        \cite{02-Mannino-Security-In-The-Land-Of-Microservices}     
            & 2017        
            & grey       
            & video         \\
        \cite{63-mateus-coehlo-security-microservices}
            & 2020
            & white
            & conference   \\
        \cite{21-Matteson-10-tips-for-securing-microservice-architecture}            
            & 2017
            & grey
            & blog post     \\
        \cite{24-Matteson-Establish-strong-microservices}
            & 2017          
            & grey            
            & blog post     \\
        \cite{26-McLarty-Securing-Microservices-APIs}          
            & 2018       
            & grey          
            & book          \\
        \cite{35-mody-zero-trust}
            & 2020
            & grey
            & blog post     \\
        \cite{56-nehme-fine-grained-access-control}
            & 2018
            & white
            & conference    \\
        \cite{43-nehme-securing-microservices}
            & 2019
            & white
            & journal       \\
        \cite{50-newman-building-microservices}
            & 2015
            & grey
            & book          \\
        \cite{13-Newman-Security-and-Microservices-by-Sam-Newman}         
            & 2016       
            & grey         
            & video         \\
        \cite{55-nkomo-software-development-microservices}
            & 2019
            & white
            & conference    \\
        \cite{58-oneill-microservice-security}
            & 2020
            & grey
            & blog post     \\
        \cite{36-parecki-oauth}
            & 2019
            & grey
            & video         \\
        \cite{39-perera-walking-the-wire}
            & 2016
            & grey
            & blog post     \\
        \hline
    \end{tabular}
\end{minipage}
\begin{minipage}{.33\textwidth}
    \centering
    \begin{tabular}{c@{\hspace{.3cm}}c@{\hspace{.3cm}}c@{\hspace{.3cm}}c}
        \hline
        \textbf{Ref.} & \textbf{Year} & \textbf{Colour} & \textbf{Type} \\
        \hline
        \cite{22-Radware-Microservice-Architectures-Challenge-Traditional-Security}          
            & 2020         
            & grey            
            & blog post     \\
        \cite{11-Raible-Security-Patterns-for-Microservice-Architectures}      
            & 2020     
            & grey         
            & blog post     \\
        \cite{62-rajasekhariah-cloud-based-microservices}
            & 2020
            & white           
            & book          \\
        \cite{54-richter-security-microservices}
            & 2018
            & white
            & conference    \\
        \cite{06-Sahni-Best-Practices-for-Building-a-Microservice-Architecture}   
            & 2019      
            & grey        
            & blog post     \\
        \cite{33-sass-security-microservices}
            & 2017
            & grey
            & blog post     \\
        \cite{47-sharma-mastering-microservices}
            & 2019
            & grey
            & book          \\
        \cite{16-Siriwardena-Microservices-Security-Landscape}       
            & 2019     
            & grey         
            & video         \\
        \cite{61-siriwardena-challenges-securing-microservices}
            & 2020
            & grey
            & blog post     \\
        \cite{25-Siriwardena-Microservices-Security-Action}
            & 2020
            & grey 
            & book          \\
        \cite{42-smith-secure-microservices}
            & 2017
            & grey
            & blog post     \\
        \cite{37-smith-secure-apis}         
            & 2019      
            & grey       
            & blog post     \\
        \cite{04-Sumo-Logic-Improving-Security-in-Your-Microservices-Architecture}      
            & 2019       
            & grey       
            & blog post     \\
        \cite{01-troisi-best-practices-microservices-app-security}        
            & 2017        
            & grey    
            & blog post     \\
        \cite{31-wallarm-shift-left}
            & 2019
            & grey
            & blog post     \\
        \cite{49-wolff-microservices-architecture}
            & 2016
            & grey
            & book          \\
        \cite{44-yarigina-security-challenges}
            & 2018
            & white     
            & conference    \\
        \cite{52-ziade-microservices-python}
            & 2017
            & grey
            & book          \\
        \hline
            \\
            \\
    \end{tabular}
\end{minipage}
\end{table}

%% file: src/3-smells-refactorings.tex
\begin{figure}
  \centering 
  \footnotesize
  \includegraphics[width=.9\textwidth]{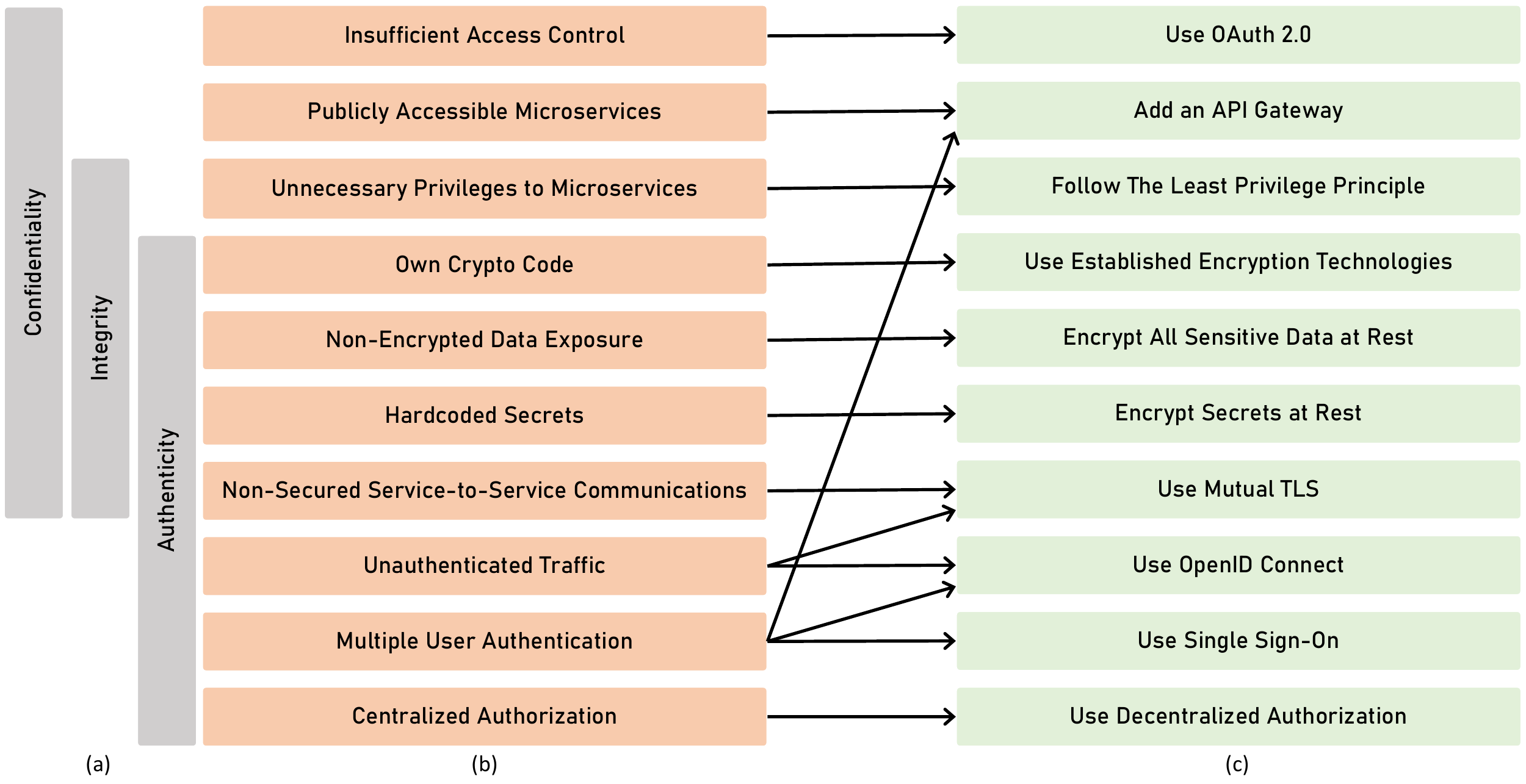}
  \caption{Taxonomy of microservices security (a) properties, (b) smells, and (c) refactorings. For the sake of readability, the association between security properties and smells is represented by aligning the corresponding boxes, whilst that between smells and refactorings is represented with arrows.}
  \label{fig:taxonomy-smells-refactorings}
\end{figure}

\Cref{fig:taxonomy-smells-refactorings} illustrates a taxonomy for the security smells pertaining to the considered security properties, and for the refactorings allowing to mitigate such smells. We obtained our taxonomy by following the guidelines for conducting systematic reviews in software engineering proposed by Petersen et al. \cite{Petersen2008}:

\begin{enumerate}
    \item We identified the security smells by performing a first scan of the selected studies.
    
    \item We excerpted the refactorings directly from the selected studies after additional scans.
\end{enumerate}

The obtained security smells and refactorings were then manually organized to obtained a taxonomy. 
A first version of the taxonomy was obtained by grouping the security smells based on the security properties they pertain to, by taking the security properties defined in the ISO/IEC 25010 standard \cite{ISO-standard} as a reference.
Out of the five properties defined in the ISO/IEC 25010, only three of them resulted to be directly corresponding to some of the identified security smells.  
These are confidentiality (\viz the degree to which a product or system ensures that data are accessible only to those authorized to have access), integrity (\viz the degree to which a system, product, or component prevents unauthorized modification of computer programs or data), and authenticity (\viz the degree to which the identity of a subject or resource can be proved to be the one claimed).
The taxonomy of security properties, smells, and refactorings underwent various iterations among the authors of this study. 
This resulted in some corrections and amendments to the first version of the taxonomy, which resulted in the final version of the taxonomy displayed in \Cref{fig:taxonomy-smells-refactorings}. 
In the taxonomy, the refactorings associated to a smell should \textit{all} be applied to mitigate its effects.

\begin{figure}
    \centering
    \includegraphics[width=.8\textwidth,trim=0 .58cm 0 .5cm]{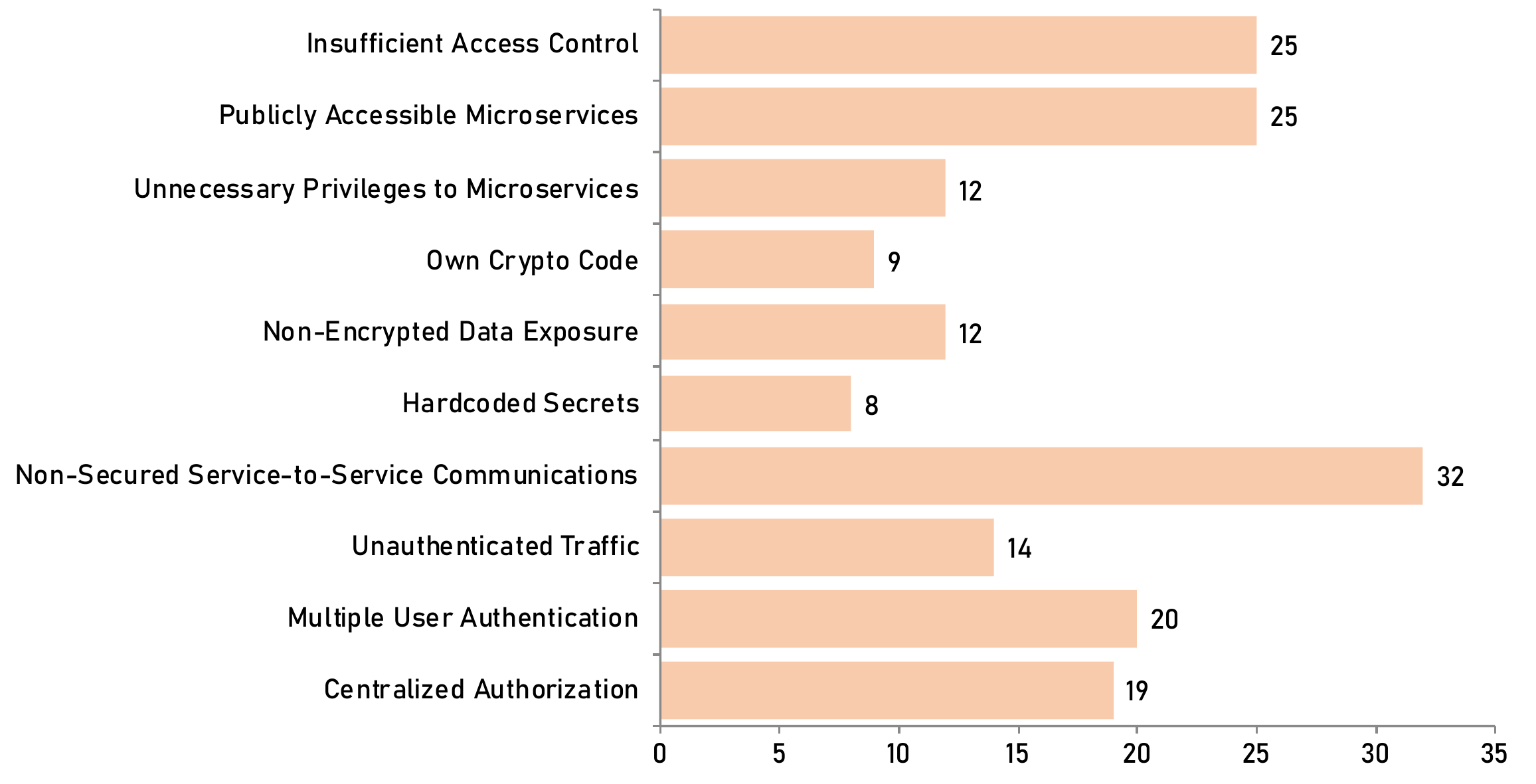}
    \caption{Coverage of the microservices security smells in the selected studies.}
    \label{fig:coverage-smells}
\end{figure}

\Cref{tab:classification} shows the classification of all selected studies based on the taxonomy in \Cref{fig:taxonomy-smells-refactorings}. 
The table provides a first overview of the coverage of the security smells over the selected studies. Such coverage is also displayed in \Cref{fig:coverage-smells}, from which we can observe that all security smells in the taxonomy are significantly recognized by the authors of the selected studies, hence making it worthy to discuss them in detail.
We hereafter discuss each identified security smell, by also highlighting how (as per what emerges from the selected studies) such smell may result in violating the corresponding security properties, as well as how each smell can be mitigated by applying a corresponding refactoring. 

\input{tables/studies-and-smells}

\input{src/3-1-confidentiality}

\input{src/3-2-integrity}

\input{src/3-3-authenticity}

%% file: tables/studies-and-smells.tex
\begin{table}
\centering
\caption{Classification of the selected studies according to the taxonomy in \Cref{fig:taxonomy-smells-refactorings}.}
\label{tab:classification}
\resizebox{\textwidth}{!}{%
\begin{tabular}{|l|c|c|c|c|c|c|c|c|c|c|}
\hline
 & \begin{tabular}[c]{@{}c@{}}Insufficient \\ access \\ control \end{tabular} & \begin{tabular}[c]{@{}c@{}}publicly \\ accessible \\ microservices\end{tabular} & \begin{tabular}[c]{@{}c@{}}unnecessary \\ privileges \\ to microservices \end{tabular} & \begin{tabular}[c]{@{}c@{}}own\\ crypto code \end{tabular} & \begin{tabular}[c]{@{}c@{}}non-encrypted\\ data\\ exposure \end{tabular} & \begin{tabular}[c]{@{}c@{}}hardcoded\\ secrets \end{tabular} & \begin{tabular}[c]{@{}c@{}}non-secured\\ service to service\\ communications \end{tabular} & \begin{tabular}[c]{@{}c@{}}unauthenticated\\ traffic \end{tabular} & \begin{tabular}[c]{@{}c@{}}multiple\\ user\\ authentication \end{tabular} & \begin{tabular}[c]{@{}c@{}}Centralised\\ authorization \end{tabular} \\ \hline
\cite{27-Abasi-Securing-modern-API-microservice} & \checkmark & \checkmark & \checkmark &  &  &  &  & \checkmark &  &  \\ \hline
\cite{34-behrens-starting-avalanche} &  &  & \checkmark &  &  &  &  &  & \checkmark &  \\ \hline
\cite{07-Boersma-Top-10-security-traps-to-avoid} &  &  & \checkmark &  & \checkmark &  & \checkmark & \checkmark &  &  \\ \hline
\cite{51-bogner-microservices-industry} &  & \checkmark &  &  &  &  &  &  &  &  \\ \hline
\cite{30-budko-5-things-api-security} &  &  &  &  &  &  &  & \checkmark &  &  \\ \hline
\cite{08-Carnell-Securing-your-microservices} & \checkmark & \checkmark & \checkmark &  &  &  & \checkmark &  &  &  \\ \hline
\cite{10-Chandramouli-Security-Strategies-for-Microservices} &  &  &  &  &  &  & \checkmark & \checkmark & \checkmark &  \\ \hline
\cite{20-da-Silva-Best-Practices-to-Protect-Your-Microservices-Architecture} &  &  & \checkmark & \checkmark & \checkmark &  & \checkmark &  & \checkmark &  \\ \hline
\cite{28-Doerfeld-Control-user-identity-microservices} & \checkmark &  &  &  &  &  &  & \checkmark &  &  \\ \hline
\cite{32-douglas-microservice-authentication} &  & \checkmark &  &  &  &  & \checkmark &  & \checkmark & \checkmark \\ \hline
\cite{12-Edureka-Best-Practices-To-Secure-Microservices} & \checkmark &  &  &  &  &  & \checkmark &  & \checkmark &  \\ \hline
\cite{45-esposito-challenges-cloud-microservices} &  &  &  &  &  &  & \checkmark &  &  &  \\ \hline
\cite{18-Feitosa-Microservice-Patterns-and-Best-Practices} &  & \checkmark &  &  &  &  & \checkmark &  & \checkmark &  \\ \hline
\cite{03-Gardner-Security-in-the-Microservices-Paradigm} &  & \checkmark &  & \checkmark &  &  &  &  & \checkmark &  \\ \hline
\cite{38-gebel-securing-apis-microservices} & \checkmark & \checkmark &  &  &  &  &  & \checkmark & \checkmark &  \\ \hline
\cite{29-Gupta-security-strategies-devops} &  &  &  &  & \checkmark &  & \checkmark &  &  &  \\ \hline
\cite{17-Hofmann-Microservices-Best-Practices-for-Java} & \checkmark &  &  & \checkmark & \checkmark & \checkmark &  & \checkmark & \checkmark & \checkmark \\ \hline
\cite{46-indrasiri-microservices-security-fundamentals} & \checkmark & \checkmark &  &  &  &  & \checkmark &  & \checkmark & \checkmark \\ \hline
\cite{48-jackson-go-microservices} &  & \checkmark & \checkmark &  & \checkmark &  & \checkmark &  & \checkmark & \checkmark \\ \hline
\cite{15-Jain-Top-10-Security-Best-Practices} &  &  & \checkmark &  & \checkmark & \checkmark & \checkmark &  &  &  \\ \hline
\cite{60-kamaruzzaman-microservice-architecture} &  & \checkmark &  &  &  &  &  &  &  &  \\ \hline
\cite{23-Kanjilal-4-fundamental-microservices-security-best-practices} &  & \checkmark & \checkmark &  &  &  & \checkmark &  & \checkmark &  \\ \hline
\cite{19-Khan-How-to-Secure-Your-Microservices} & \checkmark & \checkmark &  & \checkmark &  & \checkmark &  &  &  & \checkmark \\ \hline
\cite{05-Krishnamurthy-Transition-to-Microservice-Architecture} & \checkmark & \checkmark &  &  & \checkmark &  &  &  &  &  \\ \hline
\cite{53-lea-microservice-security} & \checkmark &  & \checkmark &  &  &  & \checkmark & \checkmark &  &  \\ \hline
\cite{40-lemos-app-security} &  &  &  & \checkmark &  &  & \checkmark &  &  &  \\ \hline
\cite{02-Mannino-Security-In-The-Land-Of-Microservices} &  & \checkmark &  &  &  & \checkmark &  &  & \checkmark & \checkmark \\ \hline
\cite{63-mateus-coehlo-security-microservices} &  &  &  &  & \checkmark &  & \checkmark &  & \checkmark &  \\ \hline
\cite{21-Matteson-10-tips-for-securing-microservice-architecture} &  &  & \checkmark &  &  &  & \checkmark &  &  &  \\ \hline
\cite{24-Matteson-Establish-strong-microservices} & \checkmark & \checkmark & \checkmark &  &  &  & \checkmark &  &  &  \\ \hline
\cite{26-McLarty-Securing-Microservices-APIs} & \checkmark & \checkmark &  &  &  &  & \checkmark & \checkmark &  & \checkmark \\ \hline
\cite{35-mody-zero-trust} &  &  & \checkmark &  &  &  & \checkmark &  &  &  \\ \hline
\cite{56-nehme-fine-grained-access-control} & \checkmark & \checkmark &  &  &  &  &  & \checkmark & \checkmark & \checkmark \\ \hline
\cite{43-nehme-securing-microservices} & \checkmark & \checkmark &  &  &  &  &  &  & \checkmark & \checkmark \\ \hline
\cite{50-newman-building-microservices} &  &  &  & \checkmark & \checkmark &  &  &  &  & \checkmark \\ \hline
\cite{13-Newman-Security-and-Microservices-by-Sam-Newman} & \checkmark &  &  &  & \checkmark &  & \checkmark &  &  &  \\ \hline
\cite{55-nkomo-software-development-microservices} &  &  &  &  &  &  & \checkmark & \checkmark & \checkmark & \checkmark \\ \hline
\cite{58-oneill-microservice-security} & \checkmark & \checkmark &  & \checkmark &  &  &  &  &  &  \\ \hline
\cite{36-parecki-oauth} & \checkmark &  &  &  &  & \checkmark &  &  &  &  \\ \hline
\cite{39-perera-walking-the-wire} &  &  &  &  &  &  &  &  &  & \checkmark \\ \hline
\cite{22-Radware-Microservice-Architectures-Challenge-Traditional-Security} &  &  &  &  &  &  & \checkmark &  &  &  \\ \hline
\cite{59-raible-11-patterns-security-microservices} & \checkmark &  &  &  &  & \checkmark & \checkmark &  &  &  \\ \hline
\cite{62-rajasekhariah-cloud-based-microservices} &  & \checkmark &  &  &  &  & \checkmark &  &  & \checkmark \\ \hline
\cite{54-richter-security-microservices} &  &  &  &  &  &  &  &  &  & \checkmark \\ \hline
\cite{06-Sahni-Best-Practices-for-Building-a-Microservice-Architecture} &  &  &  & \checkmark &  & \checkmark &  & \checkmark &  &  \\ \hline
\cite{33-sass-security-microservices} &  &  &  &  &  & \checkmark & \checkmark &  &  &  \\ \hline
\cite{47-sharma-mastering-microservices} & \checkmark &  &  &  &  &  & \checkmark &  &  &  \\ \hline
\cite{16-Siriwardena-Microservices-Security-Landscape} &  & \checkmark &  &  &  &  & \checkmark &  &  &  \\ \hline
\cite{61-siriwardena-challenges-securing-microservices} &  &  &  &  &  &  &  &  &  & \checkmark \\ \hline
\cite{25-Siriwardena-Microservices-Security-Action} & \checkmark & \checkmark &  &  & \checkmark &  & \checkmark &  & \checkmark & \checkmark \\ \hline
\cite{42-smith-secure-microservices} &  & \checkmark &  &  &  &  &  &  & \checkmark &  \\ \hline
\cite{37-smith-secure-apis} & \checkmark &  &  &  & \checkmark &  & \checkmark & \checkmark & \checkmark &  \\ \hline
\cite{04-Sumo-Logic-Improving-Security-in-Your-Microservices-Architecture} & \checkmark & \checkmark &  &  &  &  &  &  &  & \checkmark \\ \hline
\cite{01-troisi-best-practices-microservices-app-security} & \checkmark & \checkmark &  & \checkmark &  &  &  &  &  &  \\ \hline
\cite{31-wallarm-shift-left} &  &  &  &  &  &  &  & \checkmark &  &  \\ \hline
\cite{49-wolff-microservices-architecture} & \checkmark &  &  &  &  &  & \checkmark &  &  &  \\ \hline
\cite{44-yarigina-security-challenges} &  &  &  &  &  &  & \checkmark &  &  & \checkmark \\ \hline
\cite{52-ziade-microservices-python} & \checkmark &  &  &  &  &  &  &  &  & \checkmark \\ 
\hline
\end{tabular}%
}
\end{table}

%% file: src/3-1-confidentiality.tex
\subsection{Insufficient Access Control}
The \textsc{Insufficient Access Control} smell occurs whenever a microservice-based application does not enact access control in one or more of its microservices, hence possibly violating the \textsc{Confidentiality} of the data and business functions of the microservices where access control is lacking \cite{01-troisi-best-practices-microservices-app-security,08-Carnell-Securing-your-microservices}. 
If this smell is present, microservices can get exposed to, \eg the \enquote{confused deputy problem}, with attackers that can trick a service and get data that they should not be able to get \cite{13-Newman-Security-and-Microservices-by-Sam-Newman}. 
At the same time, microservices are not suitable for traditional identity control models, since client details and permissions need to be verified as and when a request is sent \cite{12-Edureka-Best-Practices-To-Secure-Microservices}, and they need a way to automatically decide whether to allow or reject calls between services \cite{52-ziade-microservices-python}. 
In addition, microservices require development teams to establish and maintain the identity of users without introducing extra latency and contention with frequent calls to a centralized service \cite{17-Hofmann-Microservices-Best-Practices-for-Java}. 

From the 25 studies describing the \textsc{Insufficient Access Control} smell, it turns out that the effects of this smell can be mitigated if development teams \textsc{Use OAuth 2.0}. 
Open Authorization (OAuth) 2.0 \cite{OAuth2} is indeed the most used mechanism to manage access delegation. 
OAuth 2.0 is a token-based security framework for delegated access control that lets a resource owner grant a client access to a certain resource on their behalf. 
This access is for a limited time and with limited scope \cite{26-McLarty-Securing-Microservices-APIs}.
OAuth 2.0 is hence a natural candidate to enforce access control in microservice-based application at each level, therein included controlling the accesses to each microservice \cite{40-lemos-app-security}.

\subsection{Publicly Accessible Microservices}
\label{sec:smells:publicly-accessible-microservices}
The \textsc{Publicly Accessible Microservices} smell occurs whenever the microservices forming an application are directly accessible by external clients \cite{24-Matteson-Establish-strong-microservices,58-oneill-microservice-security}. 
Each microservice can be accessed independently through its own API, and it needs a mechanism to ensure that each request is authenticated and authorized to access the set of functions requested \cite{27-Abasi-Securing-modern-API-microservice}. 
However, if each microservice performs authentication individually, the full set of a user’s credentials is required each time, increasing the likelihood of \textsc{Confidentiality} violations (\eg with the exposure of long-term credentials) and reducing the overall maintainability and usability of the application. 
Also, each microservice is required to enforce the security policies that are applicable across all functions of the microservice-based application \cite{26-McLarty-Securing-Microservices-APIs}.

The \textsc{Publicly Accessible Microservices} smell is discussed in 25 out of the 58 selected studies.
The authors of such 25 studies highlight that development teams can mitigate the effects of the \textsc{Publicly Accessible Microservices} smell if they \textsc{Add an API Gateway}.
This enables to identify a set of microservices to be exposed through the newly introduced API Gateway, while the rest of the microservices are made unreachable from outside this domain. 
The API gateway centrally enforces security for all the requests entering the microservices application, including authentication, authorization, throttling, and message content validation for known security threats \cite{25-Siriwardena-Microservices-Security-Action}. 
For instance, as we shall discuss in \Cref{sec:smells:multiple-user-authentication}, the API gateway can be used to implement a single sign-on to the application.
In addition, by using this approach development teams can also secure all microservices behind a firewall, allowing the API gateway to handle external requests and then communicate with the microservices behind the firewall \cite{58-oneill-microservice-security}. 
Since the clients do not directly access the services, they cannot exploit the services on their own \cite{23-Kanjilal-4-fundamental-microservices-security-best-practices}.

\subsection{Unnecessary Privileges to Microservices}
The \textsc{Unnecessary Privileges to Microservices} smell occurs when microservices are granted unnecessary access levels, permissions, or functionalities that are actually not needed by such microservices to deliver their business functions \cite{08-Carnell-Securing-your-microservices,27-Abasi-Securing-modern-API-microservice}. 
The \textsc{Unnecessary Privileges to Microservices} smell is described in 12 out of the 58 selected studies, all highlighting how such additional privileges given to microservices would potentially result in \textsc{Confidentiality} and \textsc{Integrity} issues.
This happens, \eg when a service can write or read data stored in databases or messages posted in messages queues, even if such databases or queues are not needed by the service to deliver its business function. 
As a result, resources are unnecessarily exposed, hence unnecessarily increasing the attack surface for \textsc{Confidentiality} and \textsc{Integrity} leaks: an intruder taking control of a service can indeed start reading or modifying all data and messages the service can access to \cite{53-lea-microservice-security}.

The authors of the studies describing the \textsc{Unnecessary Privileges to Microservices} smell also highlight that development teams can mitigate its effect if they \textsc{Follow the Least Privilege Principle}.
The latter recommends that accounts and services should have the least amount of privileges they need to suitably perform their business function \cite{48-jackson-go-microservices}. 
This principle should be used because even if a development team has ensured that the machine-to-machine communication is secured and there are appropriate safeguards with their firewall, there is always the risk that an attacker gets access to the microservice-based application \cite{07-Boersma-Top-10-security-traps-to-avoid,48-jackson-go-microservices}. 
Development teams should indeed limit service privileges, by providing each service with access to only the resources they actually need to deliver their business functionalities.

%% file: src/3-2-integrity.tex
\subsection{Own Crypto Code}
\label{own-crypto-code}
It is well-known that \textsc{Confidentiality}, \textsc{Integrity}, and \textsc{Authenticity} of data in software applications can get violated if development teams use their \textsc{Own Crypto Code}, \viz their own new encryption solutions and algorithms, unless they have been heavily tested \cite{19-Khan-How-to-Secure-Your-Microservices, 58-oneill-microservice-security}. 
The authors of nine out of the 58 selected studies emphasize that microservice-based applications are not the exception:
development teams that implement their own encryption solutions may end with improper solutions for securing microservices, which may result in possible \textsc{Confidentiality}, \textsc{Integrity}, and \textsc{Authenticity} issues.
The  use of \textsc{Own Crypto Code} may actually be even worse than not having any encryption solution at all, as it may produce a false sense of security \cite{50-newman-building-microservices}.

In all the studies describing the \textsc{Own Crypto Code} smell, the authors point out that the way to mitigate this smell is through the \textsc{Use of Established Encryption Technologies}. 
In other words, development teams should minimize the amount of encryption code written and maximize the amount of code that they can leverage from \enquote{bullet proof} libraries, which have already been heavily tested by the community \cite{41-gardner-security-microservices,58-oneill-microservice-security}. 
Development teams should also avoid the use of experimental encryption algorithms, as they may be subject to various kinds of vulnerabilities, which may be not yet known at the time of their use. 
Whatever are the programming languages used to implement the microservices forming an application, development teams always have access to reviewed and regularly patched implementations of established encryption algorithms \cite{50-newman-building-microservices}.

\subsection{Non-Encrypted Data Exposure}
\label{Non-Encrypted-Data-Exposure}
The \textsc{Non-Encrypted Data Exposure} smell occurs when a microservice-based application accidentally expose sensitive data, \eg because it was stored without any encryption in the data storage, or because the employed protection mechanisms are affected by security vulnerabilities or flaws \cite{07-Boersma-Top-10-security-traps-to-avoid,50-newman-building-microservices}.
When sensitive data is exposed, its \textsc{Confidentiality} and \textsc{Integrity} can get violated because it could be acquired or modified by an intruder who gets direct access to the microservices forming an application. 
As a result, the intruder may access to or manipulate, \eg some credentials for accessing other systems or business-critical data 
\cite{17-Hofmann-Microservices-Best-Practices-for-Java}.

From the 12 selected studies that describe the \textsc{Non-Encrypted Data Exposure} smell, it emerges that development teams can mitigate the effects of this smell if they \textsc{Encrypt all Sensitive Data at Rest}, \viz when data is not actively moving from device to device or network to network, e.g., when data is stored on a hard drive. 
The microservice-based architectural style enables development teams to separate functions from data in each of the microservices of an application \cite{13-Newman-Security-and-Microservices-by-Sam-Newman}. 
As a recommendation, all sensitive data should always be encrypted, and it should be decrypted only when it needs to be used. 
Most database management systems provide features for automatic encryption, and disk-level encryption features are available at the operating-system level \cite{25-Siriwardena-Microservices-Security-Action}. 
Application-level encryption is another option, in which the application itself encrypts the data before passing it over to the file system or a database.
Finally, if a caching technology is used and the data is encrypted in the database, then development teams need to ensure that the same level of encryption is applied to such caching technology \cite{48-jackson-go-microservices}.

At the same time, development teams should also keep in mind that encryption is a resource-intensive operation that could have a considerable impact on the application performance \cite{25-Siriwardena-Microservices-Security-Action}. 
As not all data needs the same level of security, and since in a microservice-based application it is common to have multiple data stores, development teams must perform a proper analysis to identify the critical ones and encrypt them.

\subsection{Hardcoded Secrets}
\label{Hardcoded-Secrets}
The \textsc{Hardcoded Secrets} smell occurs when a microservice of an application has hard-coded credentials in its source code, or when credentials are hardcoded in the deployment scripts for a microservice-based application, \eg as environment variables passing secrets in a Dockerfile or a Docker Compose file  \cite{02-Mannino-Security-In-The-Land-Of-Microservices, 17-Hofmann-Microservices-Best-Practices-for-Java}. 
Microservices are indeed likely have secrets to be used for communicating with authorization servers and other services. 
These secrets might be an API key, a client secret, or credentials for basic authentication \cite{11-Raible-Security-Patterns-for-Microservice-Architectures}. 
The authors of eight out of the 58 selected studies clearly state that development teams should never store sensitive keys and other information in environment variables. 
In the latter case, the \textsc{Confidentiality} and \textsc{Integrity} of secrets may get violated, as they could be accidentally exposed, \eg since exception handlers may grab and send the corresponding information to a logging platform. 
In addition, since child processes duplicate the parent’s environment on startup, child processes may be another source of exposure of secrets in application logs, or they may be the reason why secrets could be unintentionally accessed by other services.
In all such cases, we would end up with \textsc{Confidentiality} and \textsc{Integrity} leaks \cite{06-Sahni-Best-Practices-for-Building-a-Microservice-Architecture,19-Khan-How-to-Secure-Your-Microservices}.

The authors of the 9 studies describing the potential security leaks due to \textsc{Hardcoded Secrets} all agree on saying that development teams can mitigate the effects of this smell if they \textsc{Encrypt Secrets at Rest}.
This would indeed help achieving that only authorized resources have access to secrets.
In doing so, development teams should also adopt the following best practices: they should never store credentials alongside applications \cite{17-Hofmann-Microservices-Best-Practices-for-Java} or in the repositories used to host their source code \cite{59-raible-11-patterns-security-microservices}, nor they should exploit environment variables to pass secrets to applications \cite{19-Khan-How-to-Secure-Your-Microservices}.

\subsection{Non-Secured Service-to-Service Communications}
\label{Non-Secured-Service-to-Service-Communications}
This smell occurs whenever two microservices in an application interact without enacting a secure communication \cite{25-Siriwardena-Microservices-Security-Action, 63-mateus-coehlo-security-microservices}.
As microservice-based applications are highly distributed, communication interfaces and channels proliferate, hence increasing the overall application attack surface.
Each API exposed by each microservice indeed constitutes a potential attack vector that could be exploited by a malicious intruder, as it does each communication channel between any two microservices \cite{23-Kanjilal-4-fundamental-microservices-security-best-practices}.
Microservices often need to communicate with each other to perform their business functions, and ---if the communication channel is not secured--- the data transferred can be exposed to man-in-the-middle, eavesdropping, and tampering attacks. 
This could not only result in \textsc{Confidentiality} issues for service-to-service communications, as it could also break their \textsc{Integrity} and \textsc{authenticity}, \eg intruders could intercept the communication between two microservices and change the data in transit to their advantage \cite{20-da-Silva-Best-Practices-to-Protect-Your-Microservices-Architecture, 25-Siriwardena-Microservices-Security-Action}.

The authors of the 32 studies describing the \textsc{Non-Secured Service to Service Communications} all agree on saying that this smell can be mitigated with the \textsc{Use of Mutual Transport Layer Security}. 
Mutual TLS \cite{Siriwardena2014_MutualTLS} is indeed a widely-accepted solution to secure service-to-service communications, which enables encrypting the data in transit and ensuring its integrity and confidentiality.
This is going to protect the microservice-based application context from man-in-the-middle, eavesdropping, and tampering attacks providing a bidirectional encryption channel.
In addition, when Mutual TLS is used to secure communications between two microservices, each microservice can legitimately identify the microservice it is talking to, which means that microservices can authenticate each other \cite{25-Siriwardena-Microservices-Security-Action,46-indrasiri-microservices-security-fundamentals}. 
Hence, whereas there exist other solutions for encrypting data in transit and ensuring its integrity, Mutual TLS is suggested by experienced practitioners as it also helps mitigating the \textsc{Unauthenticated traffic} smell (as we will discuss next).

%% file: src/3-3-authenticity.tex
\subsection{Unauthenticated Traffic}
\label{sec:smells:unauthenticated-traffic}
The \textsc{Unauthenticated Traffic} smell occurs in a microservice-based application not only when there are unauthenticated API requests coming from external systems, but also when there are unauthenticated requests between the microservices of the application themselves \cite{07-Boersma-Top-10-security-traps-to-avoid, 27-Abasi-Securing-modern-API-microservice}.
To ensure \textsc{Authenticity} in microservice-based applications, it is critical to ensure that each microservice can authenticate each other, especially when the interactions are due to some user transactions and a microservice is passing the logged-in user context to the another microservice.
The problem here is that no information is typically shared among microservices, and the user context must be passed explicitly from one microservice to another. 
The challenge is hence to build trust between two interacting microservices, in such a way that the receiving microservice can accept the user context passed from the calling one \cite{25-Siriwardena-Microservices-Security-Action}. 
Therefore, a way is needed to verify that the user context (and, more generally, the data) passed between microservices has not been modified. 
If the traffic is not authenticated, there is actually no warranty that this is the case, and microservices are exposed to security attacks that may result in, \eg tampering with data, denial of service, or elevation of privileges \cite{55-nkomo-software-development-microservices}.

The \textsc{Unauthenticated Traffic} smells is discussed in 14 of the selected studies, whose authors highlight the refactorings to be enacted to mitigate this smell are to \textsc{Use Mutual TLS} and to \textsc{Use OpenID Connect}. 
As we already discussed how Mutual TLS \cite{Siriwardena2014_MutualTLS} enables authentication between any two interacting services, we hereafter focus on the use of OpenID Connect.
OpenID Connect is the most used mechanism to manage user authentication.
It is based on the use of an ID token, typically a JSON Web Token that contains authenticated user information, including user claims and other relevant attributes \cite{25-Siriwardena-Microservices-Security-Action}. 
The user ID token enables microservices to verify the user identity based on the authentication performed by an authorization server, as well as to obtain basic profile information about the end-user in an interoperable and REST-like manner.
OpenID Connect hence provides a distributed identity mechanism for traffic authentication, which is also recognized to be easy to use, self-contained, and easy to replicate \cite{28-Doerfeld-Control-user-identity-microservices}.

\subsection{Multiple User Authentication}
\label{sec:smells:multiple-user-authentication}
The \textsc{Multiple User Authentication} smell occurs when a microservice-based application provides multiple access points to handle user authentication \cite{20-da-Silva-Best-Practices-to-Protect-Your-Microservices-Architecture}. 
Each access point constitutes a potential attack vector that can be exploited by an intruder to authenticate as an end-user, and having multiple access points hence result in increasing the attack surface to violate \textsc{Authenticity} in a microservice-based application \cite{23-Kanjilal-4-fundamental-microservices-security-best-practices}. 
The use of multiple access points for user authentication also results in maintainability and usability issues, since user login is to be developed, maintained, and used in multiple parts of the application \cite{63-mateus-coehlo-security-microservices}. 

The \textsc{Multiple User Authentication} smell is discussed in 21 of the selected studie, whose authors point out that development teams can mitigate the effects of this smell if they \textsc{Use a Single Sign-On}. 
The single sign-on approach suggests having a single entry point to handle user authentication and to enforce security for all the user requests entering the microservice-based application \cite{23-Kanjilal-4-fundamental-microservices-security-best-practices}. 
This approach facilitates log storage and auditing tasks \cite{43-nehme-securing-microservices}, allowing the detection of abnormal situations that may occur. 
Single sign-on can actually be achieved if (i) we \textsc{Add an API Gateway} acting as a single entry point to the application, if not already there, and if (ii) we \textsc{Use OpenID Connect} to share the user context among the microservices. 
Both refactorings not only help implementing the single sign-on to mitigate the effects of the \textsc{Multiple User Authentication} smell, but also contribute mitigating the effects of other smells.
Indeed, we already discussed how the use of an API gateway helps mitigating the \textsc{Publicly Accessible Microservices} smell, as well as how OpenID Connect enables mitigating the effects of the \textsc{Unauthenticated Traffic} smell (in \Cref{sec:smells:publicly-accessible-microservices,sec:smells:unauthenticated-traffic}, respectively).

\subsection{Centralized Authorization}
In a microservice-based application, authorization can be enforced at the \enquote{edge} of the application (\eg with the API gateway), by each microservice of the application, or both. 
The \textsc{Centralized Authorization} smell occurs when the microservice-based application only handles authorization centrally, typically at the edge, while it does not enact any fine-grained authorization control at the microservices-level. 
Such a kind of centralized authorization diminishes the advantages of having a distributed solution, such as that given by microservices, and it reduces performances and efficiency, since the central authorization point tends to become a bottleneck \cite{17-Hofmann-Microservices-Best-Practices-for-Java, 25-Siriwardena-Microservices-Security-Action}. 
A \textsc{Centralized Authorization} may also result in violating \textsc{Authenticity} in microservice-based applications.
For instance, when authorization is managed only at the edge, microservices are exposed to the so-called \enquote{confused deputy problem}: they trust the gateway based on its mere identity, exposing the microservices in an application to misuse if they are compromised  \cite{56-nehme-fine-grained-access-control}.

The \textsc{Centralized Authorization} smell is discussed in 19 out of the 58 selected studies.
From such 19 studies, it emerges that development teams can refactor microservice-based application to mitigate the \textsc{Single Authorization} smell by enacting a decentralized authorization approach. 
They can indeed \textsc{Use Decentralized Authorization} by simply developing a token-based authorization mechanism.
This is achieved by transmitting an access token together with each request to a microservice, and access to such microservice is granted to the caller only if a known and correct token is passed \cite{54-richter-security-microservices}. 
In this way, authorization can be enforced also at the microservices-level, as it gives  each microservice more control to enforce its own access-control policies. 
It is generally recommended to have a dedicated service that acts as an authorization server, this provides three main benefits: decoupling and isolation in case the microservice-based application is compromised, separation of concerns, and an auditing point \cite{43-nehme-securing-microservices}. 
Among the 19 studies discussing the \textsc{Use Decentralized Authorization} refactoring, JSON Web Token (JWT) is the most used mechanism to implement such a refactoring. 
A JWT is a standard for safely passing claims or data attributed to a user within an environment \cite{48-jackson-go-microservices}, which ensures that a man in the middle cannot change its content because the issuer of the JWT actually signs it \cite{25-Siriwardena-Microservices-Security-Action}.

%% file: src/4-ttv.tex
Wohlin et al. \cite{Wohlin00_ExperimentationSWEngineering} define the potential threats to the validity of studies in empirical software engineering, four of which also apply to our study.
These are the threats to the \textit{external}, \textit{internal}, \textit{construct}, and \textit{conclusions} validity, which we discuss hereafter.

\smallskip \noindent
\textit{External Validity}. 
The external validity concerns the applicability of a set of results in a more general context \cite{Wohlin00_ExperimentationSWEngineering}.
Since the primary studies considered by our multivocal review were selected from a very large extent of online sources, the security smells and the corresponding refactorings may only be partly applicable to the broad area of disciplines and practices on microservices.
This may hence result in threatening the external validity of our study.

To reinforce the external validity of our findings, we organised multiple feedback sessions during our analysis of the selected studies.
The discussions held within and after the feedback sessions resulted in qualitative data, which we exploited to fine-tune both our research methods and the applicability of our findings.
We also prepared a GitHub repository\footnote{\url{https://github.com/ms-security/smells-replication-package}.} storing the artifacts produced during our study, to make them publicly available to all who wish to deepen their understanding on the data we produced.
We believe that this helps in making our results and observations more explicit and applicable in practice.

As for the external validity of our study, one may also consider our selection criteria as \enquote{too restrictive}.
However, such criteria enable focusing only on representative studies, \viz on studies that discuss at least a security smell or a corresponding refactoring.
There is however a potential risk of having missed some relevant literature, as a study might not explicitly mention the security smells in our taxonomy (\Cref{fig:taxonomy-smells-refactorings}).
To mitigate this potential threat, we carefully checked both our selection criteria against each candidate study.
We indeed checked whether a study was discussing the security issues connected to some security smell, and whether it was discussing the concrete changes to apply to a microservice-based application to mitigate the effect of such smell.
This enabled us to select also those studies that were not explicitly referring to a smell or refactoring in our taxonomy, but rather reporting on the corresponding security issues or to-be-applied changes.

Finally, another potential threat to the external validity of our study is having missed relevant grey literature. 
Practitioners may indeed share knowledge by exploiting a different terminology than ours, \eg a practitioner's blog post may discuss some security smells or refactorings, without explicitly mentioning the terms \enquote{smell} or \enquote{refactor}.
To mitigate this threat to validity, we included relevant synonyms in the search string, and we exploited the features natively supported by search engines, such as including related terms in string-based searches.

\smallskip \noindent
\textit{Construct and Internal Validity}.
Wholin et al. \cite{Wohlin00_ExperimentationSWEngineering} define the internal validity of studies as concerning the method employed to study and analyse data, therein included the potential types of bias involved.
They instead define the construct validity as the generalisability of the constructs under study. 

To mitigate the potential threats to the construct and internal validity of our study, we reduced biases by triangulation and inter-rater reliability assessment trials.
More precisely,  we adopted thematic coding \cite{Basit03_TematicCoding} to obtain the findings discussed in \Cref{sec:smells-refactorings}.
The selected studies were subject to annotation and labelling with the goal of identifying the security smells and refactoring emerging from the analysed text.
This process of analysis was executed in parallel over two 50\% splits of the selected studies, to ensure avoidance of observer bias.
The coders of the two splits (\viz the first two authors of this study) were then inverted and an inter-rater evaluation was enacted between the two emerging lists of security smells and refactorings.
To evaluate inter-rater reliability, we adopted the Krippendorff K$\alpha$ cofficient, which measures the agreement between two ordered lists of codes applied as part of content analysis \cite{Krippendorff2004_ContentAnalysis}.
We applied the K$\alpha$ coefficient to measure the agreement among the emerging lists of security smells and of their corresponding refactorings by the two independent observers who individually coded 100\% of the selected studies.
The result of applying K$\alpha$ to measure the agreement between the two emerging lists amounted to 89.1\% agreement, above the typical reference score of 80\%.
The other two authors of this study were then involved in cross-checking the coding, without any prior knowledge on the coding itself, to further mitigate potential biases.

In addition, the taxonomy organising the emerging lists of smells underwent various iterations among all the authors of this study to further avoid bias by triangulation.
The same process was applied to the actual classification of the selected studies and to the results of the analysis.

\smallskip \noindent
\textit{Conclusions Validity}.
The conclusions validity is defined by Wohlin et al. \cite{Wohlin00_ExperimentationSWEngineering} as concerning the degree to which the conclusions of a study are reasonably based on the available data.

To mitigate potential threats to the conclusions validity of our study, we exploited the above described inter-rater reliability assessment to limit potential biases in our observations and interpretations.
Additionally, the observations and conclusions discussed in this paper were drawn independently by the authors of this paper.
They were then discussed and double-checked against the selected studies in two joint discussion sessions. 

%% file: src/5-related.tex
Various secondary studies analyse and classify the state of the art and practice on microservices.
For instance, Pahl and Jamshidi~\cite{Pahl2016_MicroservicesSMS} elicit potential research directions on microservices, after discussing agreed and emerging concerns on microservices and positioning microservices with respect to existing cloud and container technologies.
Taibi et al.~\cite{Taibi2018_ArchitecturalPatternsMicroservices} instead report on common architectural patterns for microservices, by discussing the advantages, disadvantages, and lessons learned of each pattern.
However, neither Pahl and Jamshidi~\cite{Pahl2016_MicroservicesSMS} nor Taibi et al.~\cite{Taibi2018_ArchitecturalPatternsMicroservices} provide an overview on the smells possibly resulting in security issues in microservice-based applications, nor on the ways to mitigate their effects of such smells.

Other noteworthy examples are the systematic grey literature review by Soldani et al. \cite{Soldani2018_MicroservicesPainsGains} and the industrial surveys by Di Francesco et al.~\cite{DiFrancesco2018_MigratingTowardsMicroservices} and Ghofrani and L\"ubke~\cite{Ghofrani2018_ChallengesMicroservicesArchitecture}, which all provide an overview on the state of practice on microservices.
Soldani et al. \cite{Soldani2018_MicroservicesPainsGains} identify the technical and operational advantages and disadvantages of microservices, therein included the difficulties in securing microservice-based applications.
Di Francesco et al.~\cite{DiFrancesco2018_MigratingTowardsMicroservices} and Ghofrani and L\"ubke~\cite{Ghofrani2018_ChallengesMicroservicesArchitecture} instead illustrate the results of surveys they conducted with practitioners daily working with microservices, also resulting in distilling the advantages and disadvantages of microservices.
The studies by Soldani et al. \cite{Soldani2018_MicroservicesPainsGains}, Di Francesco et al.~\cite{DiFrancesco2018_MigratingTowardsMicroservices}, and  Ghofrani and L\"ubke~\cite{Ghofrani2018_ChallengesMicroservicesArchitecture} differ from ours in their objective: they report on challenges and advantages of microservices, whereas we focus on distilling the smells that may possibly result in security issues in microservices, as well as on how to mitigate their effects.

In this perspective, the industrial survey reported by Taibi and Lenarduzzi \cite{Taibi2018_BadSmells} is a step closer to ours, as they first explicitly defined 11 microservice-specific bad smells.
The smells defined by Taibi and Lenarduzzi \cite{Taibi2018_BadSmells} span from the design of microservice-based application to their actual development, and each smell is equipped with the best practices enabling to avoid incurring in such smell.
Our results complement those by Taibi and Lenarduzzi \cite{Taibi2018_BadSmells}, as none of the smells they define is about securing microservices.
We instead precisely distill the refactorings that enable mitigating the effects of well-known security smells in microservice-based applications.

Similar considerations apply to the studies by Bogner et al. \cite{Bogner2019_CollaborativeRepositoryServiceSmells}, Carrasco et al. \cite{Carrasco2018_MicroservicesSmells}, and Neri et al. \cite{Neri2020_MicroservicesSmellsRefactoring}, which all distill architectural smells for microservices.
Bogner et al.~\cite{Bogner2019_CollaborativeRepositoryServiceSmells} present a systematic literature review identifying and documenting architectural smells in SOA-based architectural styles, including microservices. 
Although the main focus of their review is on the broader SOA, several smells apply also to microservices.
Carrasco et al. \cite{Carrasco2018_MicroservicesSmells} and Neri et al. \cite{Neri2020_MicroservicesSmellsRefactoring} instead explicitly focus on microservices, with two multivocal reviews distilling architectural smells for microservices as well as architectural refactorings enabling to resolve such smells.
It is however worth noting that the focus of Bogner et al. \cite{Bogner2019_CollaborativeRepositoryServiceSmells}, Carrasco et al. \cite{Carrasco2018_MicroservicesSmells}, and Neri et al. \cite{Neri2020_MicroservicesSmellsRefactoring} is on ensuring that the architecture of microservice-based applications complies with the design principles defining microservices themselves.
We hence complement the results of their studies, as we focus on another architectural aspect than complying with microservices' design principles, \viz securing microservice-based applications. 

Finally, it is worth relating our study with the multivocal literature review by Pereira-Vale et al. \cite{pereira2021security}, and with the grey literature review by Mao et al. \cite{Mao2020_DevSecOps}, even if focused on DevOps rather than on microservices.
Pereira-Vale et al. \cite{pereira2021security} report the state of art and practice of the security solutions that have been proposed for microservice-based systems, and they identified the most used ones. 
Mao et al. \cite{Mao2020_DevSecOps} instead report on the state of practice of DevSecOps, by first overviewing the currently existing risks in classical DevOps practices, and by then illustrating the best practices in DevSecOps and how they enable addressing the security risks of DevOps.
The reviews by Pereira-Vale et al. \cite{pereira2021security} and by Mao et al. \cite{Mao2020_DevSecOps} differs from ours because they focus on the solutions proposed to support securing microservices, whereas we focus on distilling smells that may result in security issues in microservices. 
At the same time, the results in their reviews and ours complement each other, since, \eg the security solutions they review can be used to implement the refactorings we describe.

In summary, to the best of our knowledge, there is currently no study classifying the smells that can possibly result in security issues for microservice-based applications, nor eliciting the refactorings that enable mitigating their effects.
The latter is precisely the scope of our study, which we have presented in this paper.

%% file: src/6-conclusions.tex
We presented the results of a multivocal review focused on identifying the smells denoting possible security violations in microservice-based applications, and on the refactorings enabling to mitigate the effects of such smells.
More precisely, we presented a taxonomy organising ten security smells and ten refactorings.
The taxonomy associates each smell with the ISO/IEC 25010 \cite{ISO-standard} security properties it can violate, and with the refactorings that should all be applied to mitigate its effects.
We also provided an overview of the actual recognition of each smell in the selected literature, we discussed the effects of such smells in detail, and we showed how each corresponding refactoring enables mitigating their effects.

The results of our study can be of help to both researchers and practitioners interested in microservices.
Our study indeed complements other existing studies on smells and refactorings for microservices, \eg Carrasco et al. \cite{Carrasco2018_MicroservicesSmells} and Neri et al. \cite{Neri2020_MicroservicesSmellsRefactoring}, by covering the security aspects of microservice-based applications.
This can have a pragmatic value for practitioners, who can exploit the results of our study in their daily work on securing microservice-based applications.
It can also serve as a starting point for researchers wishing to study new solutions for securing microservices or to establish future research directions.

For future work, we plan to exploit the results of our study to develop an enhanced support for mitigating security smells in microservice-based applications.
For instance, we plan to extend existing languages and tools supporting the design of microservice-based applications (\eg $\mu$TOSCA and $\mu$Freshener \cite{Brogi2019_FresheningMicroservices}, or that by Pigazzini et al \cite{Pigazzini2020_MicroservicesSmellsDetection}) to enable detecting security smells automatically, as well as to automatically recommend the refactorings that should be enacted to mitigate the effects of identified security smells. 
We also plan to relate our work to the technical debt one would experience in microservice-based applications, if security smells would appear therein. 